\documentstyle[preprint,aps]{revtex}
\begin{document}
\newcommand{\lam}{\lambda}
\newcommand{\th}{\theta}
\newcommand{\fact}{\frac{1}{\sqrt{2}}}
\preprint{IP/BBSR/95-82}
\draft
\title {\bf COHERENT STATES FOR ISOSPECTRAL HAMILTONIANS}
\author{M. Sanjay Kumar \thanks{Electronic Address :
msku@iopb.ernet.in}
and Avinash Khare \thanks{Electronic Address :
khare@iopb.ernet.in}}
\address{Institute of Physics\\
Sachivalaya Marg, Bhubaneswar--751005, INDIA}
\maketitle

\begin{abstract}

We show that for the strictly isospectral Hamiltonians, the
corresponding coherent states are related by a unitary transformation.
As an illustration, we discuss, the example of strictly isospectral
one-dimensional harmonic oscillator Hamiltonians and the associated
coherent states.
\end{abstract}
\vspace{.3in}

\hspace{.35in} PACS number(s) : 03.65.Fd, 02.30.+b
\newpage

In recent years extensive work has been done on the various aspects
of coherent states.$^{1,2}$ The interest in coherent states is largely
due to the fact that they provide an alternative set of basis vectors
(non-orthogonal, overcomplete), they label the phase space of the
system, and in some cases (the harmonic oscillator being the best
known example) they have minimum fluctuations (allowed by the
Heisenberg uncertainty principle) in the canonically conjugate
variables, and hence are closest to the classical states. In recent
years, there has been a lot of interest in constructing coherent
states for potentials other than the harmonic oscillator, for
example the Morse potential$^{3}$, the Coulomb potential$^{4}$, etc.

Very recently, coherent states have also been constructed$^{5}$
for the family of Hamiltonians which are strictly isospectral to
the harmonic oscillator Hamiltonian$^{6,7}$. Various approaches exit
in the literature for the construction of such isospectral families,
for example the factorization method$^{8}$, the Gelfand-Levitan
method$^{9}$ and the approach of supersymmetry (SUSY) quantum
mechanics$^{10}$, and all of them are essentially equivalent$^{11-13}$.

In this letter we address the following question: given a Hamiltonian
whose coherent states are known, how does one construct coherent states
for the corresponding strictly isospectral Hamiltonians. In particular
we show that any two strictly isospectral Hamiltonians are related by
a unitary transformation and as a consequence the corresponding
coherent states are also related by the same unitary transformation.
An explicit construction of the unitary operators is  given. As an
illustration, we discuss the specific example of strictly isospectral
oscillator family, and show that a recent construction$^{5}$ of the
coherent states for this family of potentials is not satisfactory.
This is because these coherent states do not go over to the
harmonic oscillator coherent states in the appropriate limit.
In contrast, our coherent states do.

Consider the operators ($\hbar=m=1$)
$$
a\ =\ \fact\left(\frac{d}{dx}\ +\ W(x)\right)\ ,\
a^{\dag}\ =\ \fact\left(-\frac{d}{dx}\ +\ W(x)\right)\ ,
\eqno(1) $$
where $W(x)$ is an arbitrary function of $x$. It is well-known that the
Hamiltonians $a^{\dag}a$ and $aa^{\dag}$ are SUSY partners. As a
consequence, eigenvalues, eigenfunctions and S-matrices of the two
Hamiltonians are related, and if the eigenfunctions of $a^{\dag}a$ can
be solved for, then, the eigenfunctions of $aa^{\dag}$ can be obtained
in terms of those of $a^{\dag}a$.

Further it is also well-known that one can construct a family of
strictly isospectral Hamiltonians $b^{\dag}b$. Consider the operators
$$
b\ =\ \fact\left(\frac{d}{dx}\ +\ \hat{W}(x)\right)\ ,\
b^{\dag}\ =\ \fact\left(-\frac{d}{dx}\ +\ \hat{W}(x)\right)\ ,
\eqno(2) $$
such that
$$
bb^{\dag}\ =\ aa^{\dag}\ . \eqno(3) $$
The condition (3) leads to a Riccati equation which can be solved
to give
$$
\hat{W}(x)\ =\ W(x)\ +\ \phi_{\lam}(x)\ , \eqno(4) $$
where
$$
\phi_{\lam}(x)\ =\ \psi_{0}^{2}(x)\ \left[\lam\ +\
\int_{-\infty}^{x}dy\
\psi_{0}^{2}(y)\right]^{-1}\ , \eqno(5) $$
where $\lam$ is a real number not lying in the closed interval
$[-1,0]$ and $\psi_{0}(x)$ is the normalized ground state eigenfunction
of the Hamiltonian $a^{\dag}a$.

The eigenstates $|\th_{n}>$ of the strictly isospectral family of
Hamiltonians $H_{\lam}$ $=$ $b^{\dag}b+E_{0}$, ($E_{0}$ being the
lowest energy eigenvalue) can be obtained in terms of the
eigenstates $|\psi_{n}>$ of $H=$ $a^{\dag}a+E_{0}$, by noting
the fact that
$$
(b^{\dag}b)b^{\dag}a\ =\ b^{\dag}a(a^{\dag}a)\ .  \eqno(6) $$
Hence the normalized eigenstates of $H_{\lam}$ are
$$
|\th_{n}>\ =\ \frac{1}{E_{n}-E_{0}}\ b^{\dag}a|\psi_{n}>\ ,\
n=1,2, \ldots \eqno(7) $$
The ground state of $H_{\lam}$ is determined from the condition
$b|\th_{0}>=0$. Hence the normalized eigenfunctions of $H_{\lam}$
in the coordinate representation are given by
$$
\th_{0}(x)\ =\ \sqrt{\lam (\lam +1)}\left[\lam\ +\
\int_{-\infty}^{x}dy\
\psi_{0}^{2}(y)\right]^{-1}\psi_{0}(x)\ , \nonumber $$
$$
\th_{n}(x)\ =\ \psi_{n}(x)\ +\ \frac{1}{2(E_{n}-E_{0})}
\phi_{\lam}(x)\left(\frac{d}{dx}+W(x)\right)\psi_{n}(x)\ ,\
n=1,2, \ldots \eqno(8) $$

Let us now show that the Hamiltonians $H=a^{\dag}a+E_{0}$ and
$H_{\lam}=b^{\dag}b+E_{0}$ are related
by a unitary transformation. The condition (3) implies that if
one  writes
$$
b\ =\ aU^{\dag}\ ,\ b^{\dag}\ =\ Ua^{\dag}\ , \eqno(9) $$
then
$$
U^{\dag}U \ =\ 1\ . \eqno(10) $$
We now define the operators
$$
A\ =\ UaU^{\dag}\ ,\ A^{\dag}\ =\ Ua^{\dag}U^{\dag}\ , \eqno(11) $$
so that $b^{\dag}b$  $=$ $A^{\dag}A=$ $Ua^{\dag}aU^{\dag}$ and
hence we have the relation
$$
H_{\lam}\ =\ UHU^{\dag}\ . \eqno(12) $$
Further, the fact that the Hamiltonians $H$ and $H_{\lam}$ are
isospectral and diagonal in the orthonormal bases ${|\psi_{n}>}$
and ${|\th_{n}>}$ respectively implies that
$$
E_{n}\ =\ <\th_{n}|H_{\lam}|\th_{n}>\ =\ <\psi_{n}|H|\psi_{n}>\ ,\
n=0,1, \ldots \eqno(13) $$
Hence on using the relation (12) one has the equality (upto a
phase factor that can be taken to be unity), i.e.,
$$
U^{\dag}|\th_{n}>\ =\ |\psi_{n}>\ . \eqno(14) $$
The fact that the sets ${|\psi_{n}>}$ and ${|\th_{n}>}$ are
orthonormal implies that
$$
<\psi_{n}|\psi_{m}>\ =\ <\th_{n}|UU^{\dag}|\th_{m}>\ =\
\delta_{nm}\ , \eqno(15) $$
and hence
$$
UU^{\dag}\ =\ 1\ . \eqno(16) $$
Thus we have shown that the strictly isospectral Hamiltonians and
their respective eigenfunctions are related by a unitary
transformation.

One way of defining coherent states for an arbitrary Hamiltonian is
based on the dynamical symmetry group of the Hamiltonian$^{2}$.
Let $G$ be the symmetry group of the Hamiltonian $H$. Hence one
can express $H$ as a linear combination of the generators
${J_{i}}$ of the Lie algebra of $G$. Thus
$$
H\ =\ \sum_{i}\ d_{i}\ J_{i}\ , \eqno(17) $$
where $d_{i}$ are complex numbers and $J_{i}$ are closed under the
commutation relations
$$
\left[J_{i}\ ,\ J_{j}\right]\ =\ c_{ij}^{k}\ J_{k}\ . \eqno(18) $$
Now, since $H_{\lam}$ is unitarily related to $H$, one has
$$
H_{\lam}\ =\ \sum_{i}\ d_{i}\ \tilde{J}_{i}\ ,\ \tilde{J}_{i}\
=\ UJ_{i}U^{\dag}\ , \eqno(19) $$
where $\tilde{J}_{i}$ obey the same Lie algebra (18) as do
$J_{i}$, since the structure constants $c_{ij}^{k}$ do not
change under a unitary transformation.
Thus $H_{\lam}$ has the same symmetry group as that of
$H$. Let $D(z)$ be the element belonging to the coset space of $G$
with respect to its maximally stability subgroup. Note that $D(z)$ is
an operator  function of the generators $J_{i}$. If the Perelomov
coherent states associated with $H$ and $H_{\lam}$ are defined
respectively as
$$
|z>\ =\ D(z)\ |\psi_{0}>\ , \nonumber $$ $$
|z;\lam>\ =\ D_{\lam}(z)\ |\th_{0}>\ , \eqno(20) $$
then, as consequence of Eq.(19) one has
$$
D_{\lam}(z)=UD(z)U^{\dag}\ , \eqno(21) $$
and hence it follows from Eq.(14) that
$$
|z;\lam>\ =\ U\ |z>\ . \eqno(22) $$
Thus we have shown that the Perelomov coherent states associated with
the isospectral Hamiltonians $H$ and $H_{\lam}$ are unitarily
related.

The explicit structure of the unitary operator $U$ can be easily
obtained. Expanding $U$  in the complete set of eigenstates of $H$
and using Eq.(14) we have
$$
U\ =\ \sum_{n,m=0}^{\infty}\ U_{nm}\
|\psi_{n}><\psi_{m}|\ ,\ U_{nm}\ =\ <\psi_{n}|\th_{m}>\ .
\eqno(23) $$
Using Eq.(8) one can write down explicit expressions for the
matrix elements $U_{nm}$. Thus
$$
U_{n0}\ =\ \int dx\ \sqrt{\lam (\lam +1)}
\left[\lam\ +\ \int_{-\infty}^{x}dy\
\psi_{0}^{2}(y)\right]^{-1}
\psi_{n}^{*}(x)\psi_{0}(x)\ , \nonumber $$
$$
U_{n,m+1}\ =\ \delta_{n,m+1}\ +\ \frac{1}{2(E_{m+1}-E_{0})}
\int dx\psi_{n}^{*}(x)\phi_{\lam}(x)
\left(\frac{d}{dx}+W(x)\right)\psi_{m+1}(x)
\ ,\ n,m=0,1, \ldots \eqno(24) $$
It is simple to see from Eqs.(5) and (24) that in the limit
$|\lam|\rightarrow\infty$ one has $\phi_{\lam}(x)\rightarrow 0$ and
hence $U^{\dag}$, $U$ $\rightarrow 1$.  As a result
$H_{\lam}\rightarrow H$,
and hence the eigenstates as well as the coherent states associated
with $H_{\lam}$ reduce to those of $H$ in this limit.

We next consider the specific example of the strictly isospectral
oscillator Hamiltonians in order to illustrate the general arguments
made in the foregoing. The harmonic oscillator is described by the
Hamiltonian $H=a^{\dag}a+\frac{1}{2}$ $(\hbar=m=\omega=1)$ where
$$
a\ =\ \fact\left(\frac{d}{dx}\ +\ x\right)\ ,\
a^{\dag}\ =\ \fact\left(-\frac{d}{dx}\ +\ x\right)\ ,
\eqno(25) $$
where the annihilation and creation operators $a$ and $a^{\dag}$
obey the usual commutation relations $[a,a^{\dag}]=1$.
The isospectral oscillator family is then described by the
Hamiltonian $H_{\lam}$ (see Eq.(12))
$$
H_{\lam}\ =\ A^{\dag}A+\frac{1}{2}\ ,\ A\ =\ UaU^{\dag}\ ,\
A^{\dag}\ =\ Ua^{\dag}U^{\dag}\ . \eqno(26) $$
The structure of the unitary operator $U$ [see Eq.(24)]
in this case will depend on the eigenvalues and
eigenfunctions of the harmonic oscillator $H$.
Note that $A$ and $A^{\dag}$ obey the same commutation relations as
do $a$ and $a^{\dag}$, i.e., $[A,A^{\dag}]=1$. Thus $A$ and
$A^{\dag}$ are indeed the annihilation and creation operators
associated with the isospectral Hamiltonian $H_{\lam}$. Let
us define new canonically conjugate operators, namely
$$
\hat{X}\ =\ \fact(A^{\dag}+A)\ ,\
\hat{P}\ =\ \frac{i}{\sqrt{2}}(A^{\dag}-A)\ ,\
\nonumber $$ $$
[\hat{X},\hat{P}]\ =\ i\ , \eqno(27) $$
so that the Hamiltonian $H_{\lam}$ can be written in the form
$$
H_{\lam}\ =\ \frac{1}{2}(\hat{P}^{2}\ +\ \hat{X}^{2})\ .
\eqno(28) $$
Note that $\hat{X}$ and $\hat{P}$ are related to the position
and momentum operators $\hat{x}$ and $\hat{p}$ of the harmonic
oscillator by the unitary transformation as in Eq.(26).
Thus the family of isospectral oscillators can be viewed as
harmonic oscillators but expressed in terms of appropriately
transformed position and momentum operators. As a consequence,
the coherent states associated with the isospectral oscillator
family $|z;\lam>$ may be defined as eigenstates of the
annihilation operator $A$, namely
$$
A\ |z;\lam>\ =\ z\ |z;\lam>\ , \eqno(29) $$
where the eigenvalue $z$ is $\lam$-independent, or equivalently,
in the Perelomov sense as the displaced ground state, namely
$$
|z;\lam>\ =\ D_{\lam}(z)|\th_{0}>\ ,\
D_{\lam}(z)\ =\ exp\{zA^{\dag}-z^{*}A\}\ , \eqno(30) $$
or, equivalently, as the state which has the minimum uncertainty
product
$$
\Delta X\Delta P\ =\ \frac{1}{2}\ , \eqno(31) $$
(note that we have chosen $\hbar=m=\omega=1$)
with the uncertainties in $X$ and $P$ being equal. It must be
noted that these coherent states are not minimum uncertainty states
with respect to the position and momentum of the particle, viz., $x$
and $p$. As is well known, only the Gaussian states minimize the
product $\Delta x\Delta p$. As argued more generally in the
foregoing the coherent states of the isospectral oscillator
$|z;\lam>$ are related to the harmonic oscillator coherent
states by a unitary transformation as in Eq.(26).

One may also define a more general state which minimizes the
uncertainty product $\Delta X\Delta P$, in analogy with the
squeezed coherent state$^{14}$ of the usual harmonic oscillator, as
follows:
$$
|\xi,z;\lam>\ =\ S_{\lam}(\xi)D_{\lam}(z)|\th_{0}>\ ,\
S_{\lam}(\xi)\ =\ exp\{\frac{1}{2}\xi(A^{\dag})^{2}-
\frac{1}{2}\xi^{*}A^{2}\}\ , \eqno(32) $$
where the displacement operator $D_{\lam}(z)$ is as defined in
Eq.(30). Note that in the state $|\xi,z;\lam>$ the uncertainties
$\Delta X$ and $\Delta P$ are unequal while the product is one-half.
As a consequence of Eq.(26) it follows that the state $|\xi,z;\lam>$
is related to the squeezed coherent state of the harmonic oscillator
by a unitary transformation.

There has been some discussion in the literature$^{5,6}$ about
what is the correct set of annihilation and creation operators,
and the coherent states associated with the isospectral
oscillator family. For example in Ref.6, annihilation and
creation operators ($A_{M}, A_{M}^{\dag}$)
are constructed. However they do not connect the ground state
$|\th_{0}>$ to $|\th_{n}>$ ($n\geq1$). Further they do not reduce
to the oscillator operators ($a,a^{\dag}$) in the limit $|\lam|$
$\rightarrow\infty$ but instead reduce to ($a^{\dag}a^{2}$,
$(a^{\dag})^{2}a$).  In Ref.5, coherent states are constructed as the
eigenstates
of the annihilation operator $A_{M}$ of Ref.6 and these consequently
do not reduce to the harmonic oscillator coherent states in the
limit $|\lam|$ $\rightarrow\infty$. We would like to emphasize that
unlike Ref.6 our ($A,A^{\dag}$) defined by Eq.(27) are the correct
set of annihilation and creation operators for the isospectral
oscillator family, they act on ${\it all}$ the eigenstates of the
Hamiltonian $H_{\lam}$, and they reduce to ($a,a^{\dag}$) in the limit
$|\lam|$ $\rightarrow\infty$. Further the coherent states associated
with $H_{\lam}$ also reduce, unlike in Ref.5, to the harmonic
oscillator coherent states in this limit.  The new coherent states
$|z;\lam>$ possess all the properties of the usual coherent states
$|z>$ such as non-orthogonality, overcompleteness, etc., as these
properties are invariant under a unitary transformation.

In conclusion, we have demonstrated that the strictly isospectral
family of Hamiltonians are related bya unitary transformation, and
argued that, as a consequence, the coherent states associated with
these isospectral Hamiltonians are also related by the same
transformation. We have given an explicit construction of the
unitary transformation. We would like to remark that the conclusions
of this letter are valid even in the case of $n$-parameter
isospectral families of Hamiltonians$^{15}$.

MSK gratefully acknowledges discussions with Professor
V. Srinivasan. \\ \\
{\bf References:}
\begin{enumerate}
\item Klauder J R and Skagerstam B S 1985, {\it {Coherent States -
Applications in Physics and Mathematical Physics}} (Singapore:
World Scientific).
\item Perelomov A 1986, {\it {Generalized Coherent States and Their
Applications}} (Texts and Monographs in Physics) (Berlin: Springer).
\item Nieto M M and Simmons Jr. L M 1979, Phys. Rev. D{\bf 20}, 1342.
\item Bhaumik D, Dutta-Roy B and Ghosh G 1986, J. Phys. A: Math. Gen.
{\bf 19} 1355; \\
Gerry C C 1986, Phys. Rev. A{\bf 33}, 6 and references therein.
\item Fernandez D J, Hussin V and Nieto L M 1994, J. Phys. A:
Math. Gen. {\bf27}, 3547.
\item Mielnik B 1984, J. Math. Phys. {\bf 25}, 3387.
\item Khare A and Sukhatme U 1989, J. Phys. A: Math. Gen. {\bf 22},
2847.
\item Infeld L and Hull T E 1951, Rev. Mod. Phys. {\bf 23}, 21.
\item Chadan K and Sabatier P C 1977, {\it {Inverse Problems in
Quantum Scattering Theory}} (Berlin: Springer).
\item For a recent review see Cooper F, Khare A and Sukhatme U
1995, Phys. Rep.{\bf 251}, 267.
\item Sukumar C V 1985, J. Phys. A: Math. Gen. {\bf 18}, L57, 2917; \\
Cooper F, Ginocchio J N and Khare A 1987, Phys. Rev. D{\bf 36}, 2458.
\item Nieto M M 1984, Phys. Lett. {\bf 145B}, 208; \\
Cao X C 1991, J. Phys. A: Math. Gen. {\bf 24}, L1155.
\item Sukumar C V 1985, J. Phys. A: Math. Gen. {\bf 18}, 2937; \\
Chaturvedi S and Raghunathan K 1986, J. Phys. A: Math. Gen. {\bf 19},
L775.
\item Yuen H P 1976, Phys. Rev. A{\bf 13}, 2226.
\item Keung W Y, Sukhatme U, Wang Q and Imbo T 1989, J. Phys. A:
Math. Gen. {\bf 22}, 987.
\end{enumerate}
\end{document}